\documentstyle[aps,prl,preprint]{revtex}
\begin{document}

\title{The energy flux into a fluidized granular medium at a vibrating wall}

\author{Sean McNamara and Jean-Louis Barrat}
\address{ D\'epartement de Physique des Mat\'eriaux \\
Universit\'e Claude Bernard and CNRS, 69622 Villeurbanne Cedex, France}

\date{\today}

\maketitle

\begin{abstract}

We study the power input of a vibrating wall into a fluidized granular
medium, using event driven simulations of a model granular system.  The
system consists of inelastic hard disks contained between a stationary
and a vibrating elastic wall, in the absence of gravity.  Two scaling
relations for the power input are found, both involving the pressure.
The transition between the two occurs when waves generated at
the moving wall can propagate across the system.  Choosing an
appropriate waveform for the vibrating wall removes one of these
scalings and renders the second very simple.

\end{abstract}

\pacs{5.60.+w, 5.70+Ln, 46.10.+z, 62.90.+k}

One of the essential differences between fluidized granular systems and
usual gases is that sustaining a  fluidized state necessitates a
continuous input of energy into the system, since the particle
kinetic energy is dissipated during the collisions. Experimentally,
this is often achieved by using a vibrating piston. The nature of the
energy exchange between the vibrating piston and the fluidized granular
medium, however, does not appear to have been studied in great detail.
In most cases, it is assumed that the vibrating wall imposes a
``granular temperature'' of the particles that corresponds to its mean
squared velocity. The purpose of this work is to achieve  a more
detailed understanding of this energy exchange by studying numerically
and theoretically a particularly simple case. The system we consider
(figure \ref{fig:syswave}a) is a two dimensional fluid of inelastic hard
discs, contained between two walls in the $y$ direction and with
periodic boundary conditions in the $x$ direction.  The moving wall is,
at its lowest point, at $y=0$, while a stationary wall limits the
system at $y=H$. For the sake of simplicity, we have chosen to  treat
the wall/particle collisions as elastic, and to set the gravity force
equal to zero.  Hence the system can be entirely characterized by a
small number of dimensionless parameters. These parameters are the
ratios of the system sizes $H$ (in the $y$ direction) and $L$ (in the
$x$ direction) to the particle radius $a$, the density measured by the
area fraction $N \pi a^2/LH$ ($N$ is the number of particles),  the
amplitude of vibration $A$ of the moving wall, measured in units of
$a$, and the restitution coefficient $r<1$.  [In the center of mass
frame of two colliding particles $v_n'=-rv_n$, where $v_n$ ($v_n'$) is
the normal component of the particles' velocity before (after) the
collision.]  Finally, the problem is completely defined by specifying
the waveform $\phi(t)$ of the wall vibration.  Note that $\tau$,
the period of this waveform defines the time unit in the problem.
There is a second timescale in the problem: $t_{\text{coll}}$, the
time between collisions experienced by an average particle.  But
$t_{\text{coll}}$ is not independent of $\tau$; the ratio
$\tau/t_{\text{coll}}$ is a function of the five dimensionless numbers
given above.  In the simulations considered here, $2 \le \tau/t_{\text{coll}}
\le 40$.  In figure \ref{fig:syswave}b, we show the two waveforms,
labeled (A) and (B), used to drive the vibrating wall.

We note that the system studied in this paper is an externally driven
version of the system considered in references \cite{GZ,MY,DB}. Despite its 
simplicity, this system was shown to display a nontrivial behavior even in the absence of external forcing, with the development of several instabilities
during ``homogeneous cooling''.  Other instabilities, such as the formation
of lateral structures in the $x$ direction, could be expected in the forced case. Since our main object is the study of energy input 
at a local scale, we deliberately avoided such structures by 
using a relatively small system width, $L/a = 50$.

If we were to add gravity to the system studied in this paper, we would
have the system studied in references \cite{WHJ,LHB}.  These references
present discrepancies between theory and experiment which could be
resolved by insights presented in this paper.

Figure \ref{fig:wfcompare} shows that a detailed understanding of the
particle-wall interaction is needed.  When the wall is driven with the
asymmetric wave form (B), the relation between the average energy per
particle $E/N$ and the restitution coefficient obeys a simple power law
$E/N\sim(1-r)^{-1.9}$.  On the other hand, the symmetric waveform (A)
generates much more complicated behavior.  Since the only difference
between these two curves is the waveform, their differences cannot be
explained without understanding what happens at the vibrating
boundary.  This paper explains how the waveform causes the two
different relations between $E/N$ and $r$.

We begin by looking closely at what is happening inside the system.  We
show typical density and temperature profiles in figure
\ref{fig:profiles}, for a system driven by a symmetric waveform (type
(A) in figure \ref{fig:syswave}b). The evolution of the profiles
during the wall motion is also detailed in these figures. As the
vibrated system is ``heated'' by the moving wall, an inhomogeneous
density and temperature (temperature being understood here as kinetic
energy per particle) profile develops.  Far from the moving wall, the
system is denser and cooler than close to it.  The temperature profile
clearly displays two different regions. In a region that extends over
about half the height $H$ of the box, a heat pulse generated at the
vibrating wall propagates in the positive $y$ direction.  Farther away
from the moving wall, the heat pulses are completely damped and the
temperature is stationary.  The ``boundary region'' for the temperature
thus appears to be rather broad.  The density profile also displays a
(small) time dependent component, indicating that the heat pulses are
coupled to compression waves in the fluid.  These heat and density
waves can transport significant amounts of energy within the boundary
region.  This is in conflict with the assumption that energy transport
is dominated everywhere by conduction (for example in \cite{Lee1}).  We
note that similar waves have been seen in simulations of shaken
granular materials under gravity \cite{Lee2}.  These waves resemble
sound waves in a gas.  However, their description in terms of a single
``temperature'' is not perfectly accurate. A more careful examination
shows that the particles in these waves can be divided into two
distinct  populations with significantly different kinetic
temperatures. One population is made up of rapidly moving particles
who, having just encountered the moving wall, travel towards the
stationary region, carrying the heat pulses. The other is a population
of slowly moving particles emerging from the stationary region, and
traveling towards the moving wall.  

As $r$ is increased towards $1$, these pulses broaden, and they
penetrate farther into the stationary region.  Eventually, they reach
the stationary wall, so that the boundary region extends to
the whole simulation box. 

We now seek a law giving the power injected by the wall, $P_w$ in terms
of the kinetic pressure $p$  (defined as the momentum transfer to the
stationary wall per unit surface and time).  Because of momentum
conservation, the pressure on the vibrating wall must also be $p$, and
dimensional reasoning suggests that the power input should be
proportional to $p$ times the wall velocity $V$. For the asymetric
waveform  (B), this proportionality is indeed easily shown to hold as
an equality. The argument is as follows. Collisions between the
particles and the wall take place only when the wall is in its
ascending phase.  When such a collision takes place, the energy change
and the momentum change of the particle are related by $\Delta E = V
\Delta p_y$.  Summing over all particles that hit the wall during a
cycle shows that the average energy transfer per unit time will be
equal to the wall velocity multiplied by the momentum transfer per unit
time, i.e. $P_w=pVL$. This conclusion is extremely well borne out by the
simulation results, as can be seen in figure \ref{fig:scaling1}.

The reasoning can be generalized  to the case of other waveforms, e.g.
(A).  In that case, the particles can either receive or loose energy as
they hit the wall.  If the arrival times of the particles at the
vibrating wall are independent of the phase of the vibrating wall, then
the probabilities of these two events will depend only on the ratio
between velocity of the particles and the wall velocity $V$, so that we
expect the power input to scale as $pV\!L\,F(V/U) $, where $U$ is a
velocity characteristic of the particles that hit the wall, and $F$ is
a dimensionless function that will depend on the waveform and of the
velocity distribution of the particles near the wall. In figure
\ref{fig:scaling1}, this scaling relation was tested by plotting the
power input as a function of the dimensionless variable $V/U$, where
the typical particle speed $U$ is estimated by the square root of the
average energy per particle, $(E/N)^{1/2}$.  The unscaled values of
$P_w$ range over four orders of magnitude, so the success of the
scaling is impressive.  The scaling is very well obeyed except for the
largest amplitudes of wall vibration, in which case it fails for small
values of the rescaled power input $P_w/pV\!L$.

This failure of the scaling relationship can be traced back to the
establishment of the second situation mentioned above, namely that in
which the ``boundary''  region extends over the whole simulation cell.
The transition to this situation is observed for values of the
restitution coefficient very close to one and for large vibration
amplitudes, and only in the case where the excitation is of the form
(A). When this transition takes place, the points in figure
\ref{fig:scaling1} leave the scaling curve, displaying a discontinuous
and nonmonotonous behavior.  The reason for this behavior can be
understood as follows. The heat pulse and associated compression wave
now reach the upper elastic wall before disappearing. They are
reflected at this wall, and eventually hit the moving wall again. The
arrival times of the particles at the moving wall are no longer
independent of the phase of the wall vibration, so that the simple
assumptions used in deriving the scaling relationship break down.
Moreover, when the energy input takes place by such a correlated
collision mechanism, a nonmonotonic behavior can be understood as a
resonance between the travel time of the wave and the period of wall
motion. 

This physical picture suggests a second scaling relationship.  In
figure \ref{fig:scaling2}, we show that $P_w = (pV^2\tau L/H)\, G[U\tau
/(H-A)]$, where $G$ is another dimensionless function.  (The inclusion
of the period of the wall vibration $\tau$ is required dimensionally.)
This second scaling is valid everywhere the first one fails.  It can be
understood by considering the propagation of sound waves in the box.
The wave speed scales as $U$, so $U\tau/(H-A)$ is the fraction of the
box that a wave can travel during one period.  For particular values of
$U\tau/(H-A)$, resonance between the wall and the waves will occur.
$P_w$ will scale as $\hat p V\!L$, where $\hat p$ is the pressure
amplitude of the wave.  Examining the properties of sound waves in a
compressible gas at pressure $p$, we find that the pressure and
velocity amplitudes are related by $\hat p = (k/\omega) p \hat u$,
where $\hat u$ is the velocity amplitude, $k$ is the wavenumber of the
wave, and $\omega$ is its frequency.  Setting $\hat u \sim V$, $k\sim
H^{-1}$ and $\omega\sim \tau^{-1}$ gives the scaling in figure
\ref{fig:scaling2}.

The resonance affects the power injected by the wall only for the
symmetric waveform (A), even though waves generated by the asymmetric
waveform (B) can also propagate throughout the box at large $A$ and $r$
close to $1$.  The reason is that particles can either gain or loose
energy with the symmetric waveform.  Thus, shifting the arrival time of
a large group of particles by half a period can change the sign of
$P_w$.  On the other hand, for the asymetric waveform (B), the amount
of energy gained by the particles does not depend on the phase of the
wall.

The transition between the two scalings occurs at the critical value
$U\tau/(H-A)\approx 0.4$.  Examination of simulations made with $30<H<100$
confirms that this critical value remains constant.  At this time, we
do not have a an explanation for this critical value, nor a detailed
understanding of the transition.

We believe these results to be relevant to current experimental
questions.  First of all, an experimental version of this system will
soon be studied in microgravity \cite{Fauve}.  Second, these results
can easily be extended to experiments done in gravity by realizing that
conservation of momentum requires that the pressure (the time-averaged
force on the bottom plate) be the weight of the granular material:
$pL=Nmg$.  Finally, this work suggests that using the waveform (B) [or
an experimental approximation] may simplify results, leading to a
better physical understanding of granular flows.

% \acknowledgments 
This work was supported by the 
Centre National d'Etudes Spatiales. S.~McNamara 
benefited from a Region Rhones-Alpes visiting scientist position
at the Pole Scientifique de Mod\'elisation Num\'erique of ENS-Lyon.

\begin{figure}
\caption{(a) A sketch of the simulated system.  (b)
The two waveforms used to drive the vibrating wall: the symmetric
waveform (A) and the asymmetric waveform (B).}
\label{fig:syswave}
\end{figure}

\begin{figure}
\caption{The average energy per particle as a function of $r$, showing
the effect of changing the waveform of the vibrating wall.  The parameters
of these simulations are $L=H=50a$, area fraction $N\pi a^2/LH=0.25$,
wall velocity $V=8a/\tau$, and $0.8 \leq r \leq 0.998866$.}
\label{fig:wfcompare}
\end{figure}

\begin{figure}
\caption{Profiles of temperature (a) and density (b), (measured by the
local area fraction $\nu$).  The nondimensional paramters are
$L=H=50a$, $r=0.95$, $N\pi a^2/LH=0.25$, $A=2$ and the symmetric
waveform.  [There are about $N/(L/2a)\approx8$ layers of particles.] In
each graph, there are four lines showing the field values at four times
during the driving cycle.  The wall is at its lowest point ($y=0$) at
$t=0$, and at its highest point ($y=A=2$) at $t=0.5$.  At $t=0.25$
($t=0.75$), it is halfway between these extremes, and ascending
(descending).}
\label{fig:profiles}
\end{figure}

\begin{figure}
\caption{The power input scaled as $P_w = pV\!L\, F(U/V)$.  The bold points
were generated by the symmetric waveform; the ligher points by the
symmetric.  The paramters are $L=H=50a$, $N\pi a^2 = 0.25$, $1 \leq A \leq 5$
(as indicated on the graph), and $0.8 \leq r \leq 0.998866$.}
\label{fig:scaling1}
\end{figure}

\begin{figure}
\caption{The same data as the bold points in figure \ref{fig:scaling1}
(symmetric waveform), but scaled with the second scaling presented in
the text: $P_w= (pV^2L\tau/H)\, G[U\tau/(H-A)]$.  The points which
disobey the previous scaling collapse onto a single curve.  The gap
at $0.5 \leq U/(H-A) \leq 0.8$ is caused by the resonance between
the sound waves and the vibrating wall.  This gap corresponds to the
discontinuities in figures \ref{fig:wfcompare} and \ref{fig:scaling1}.}
\label{fig:scaling2}
\end{figure}

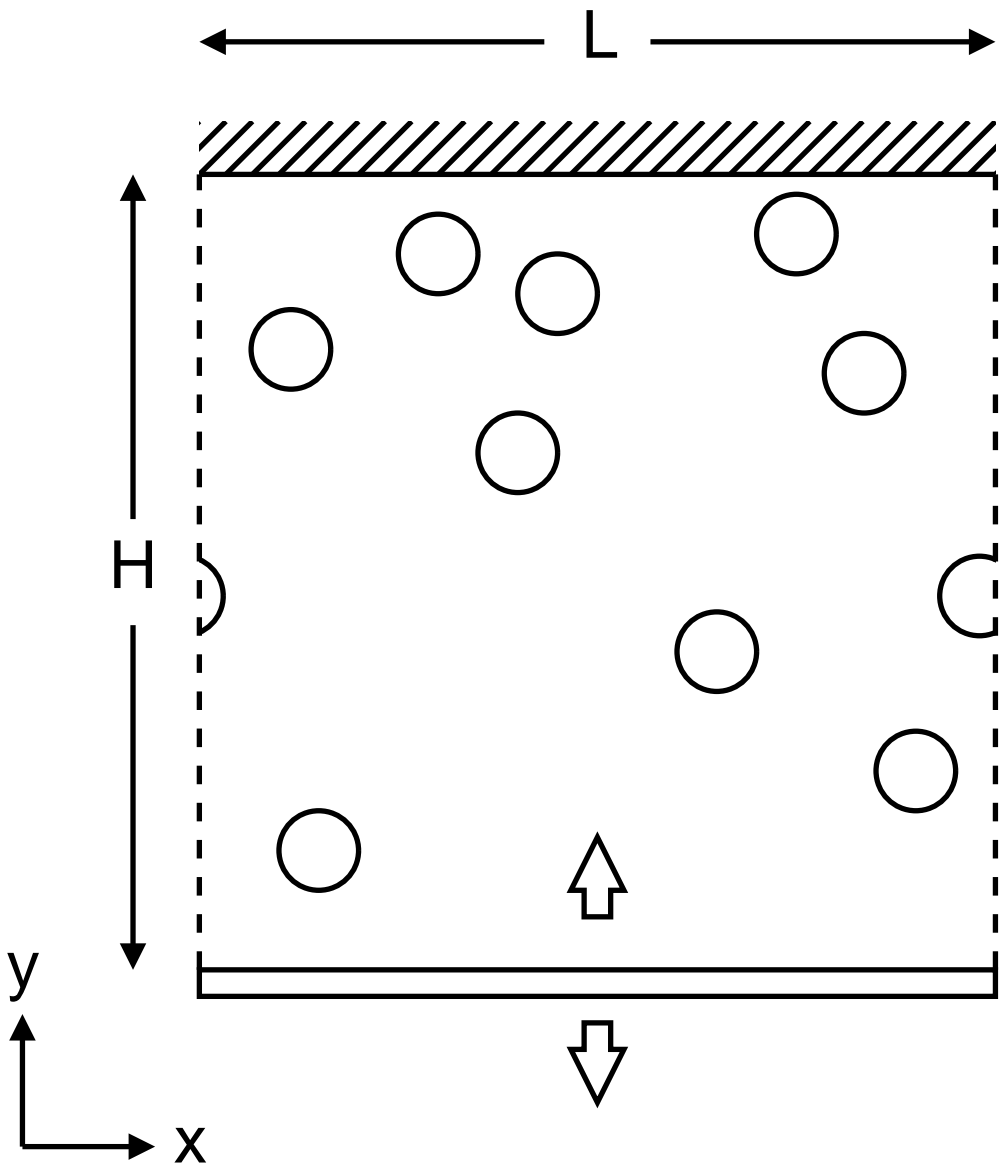
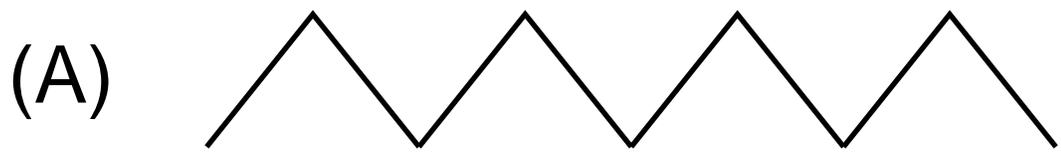
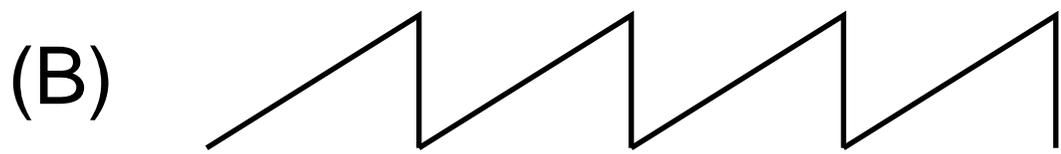
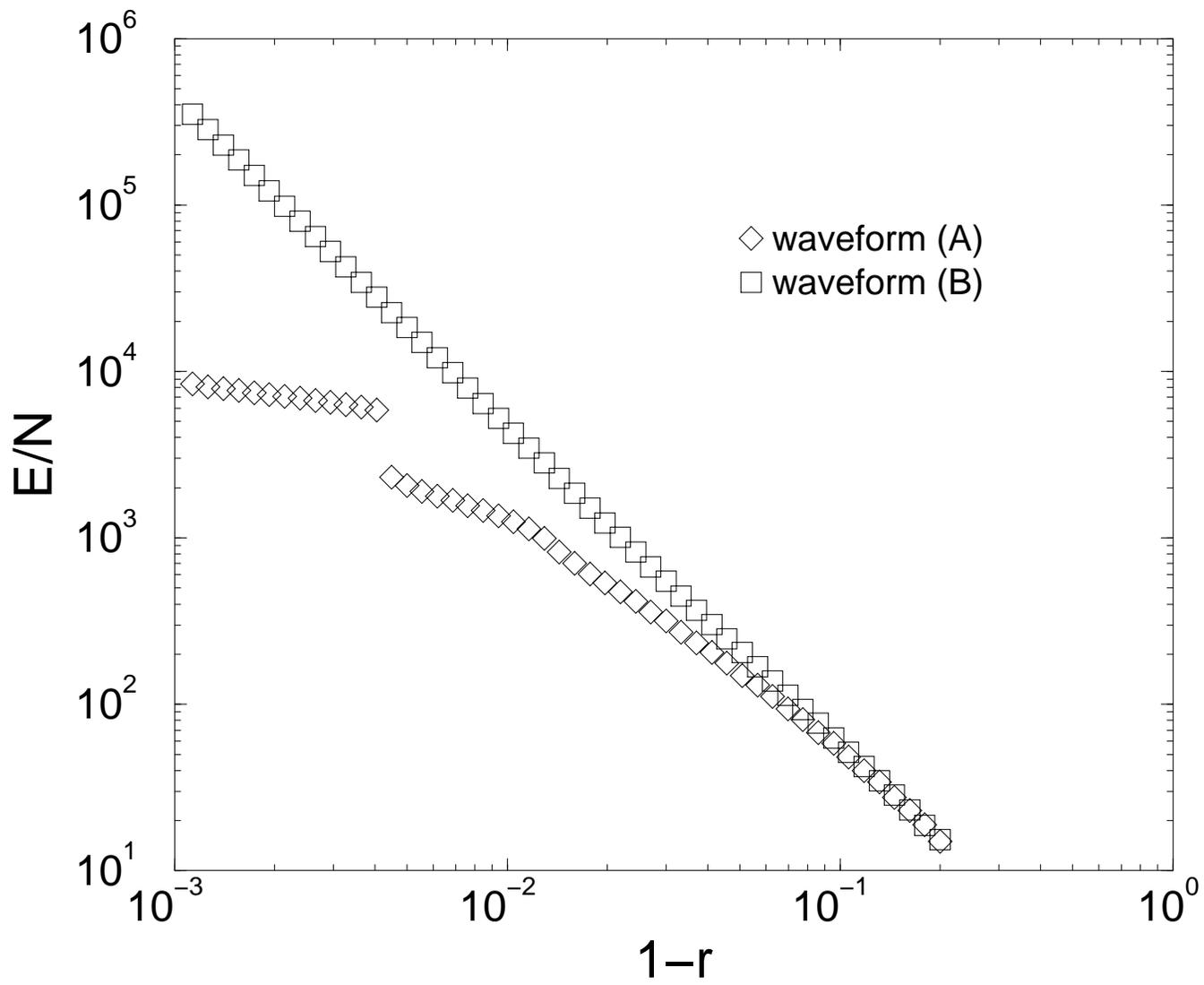
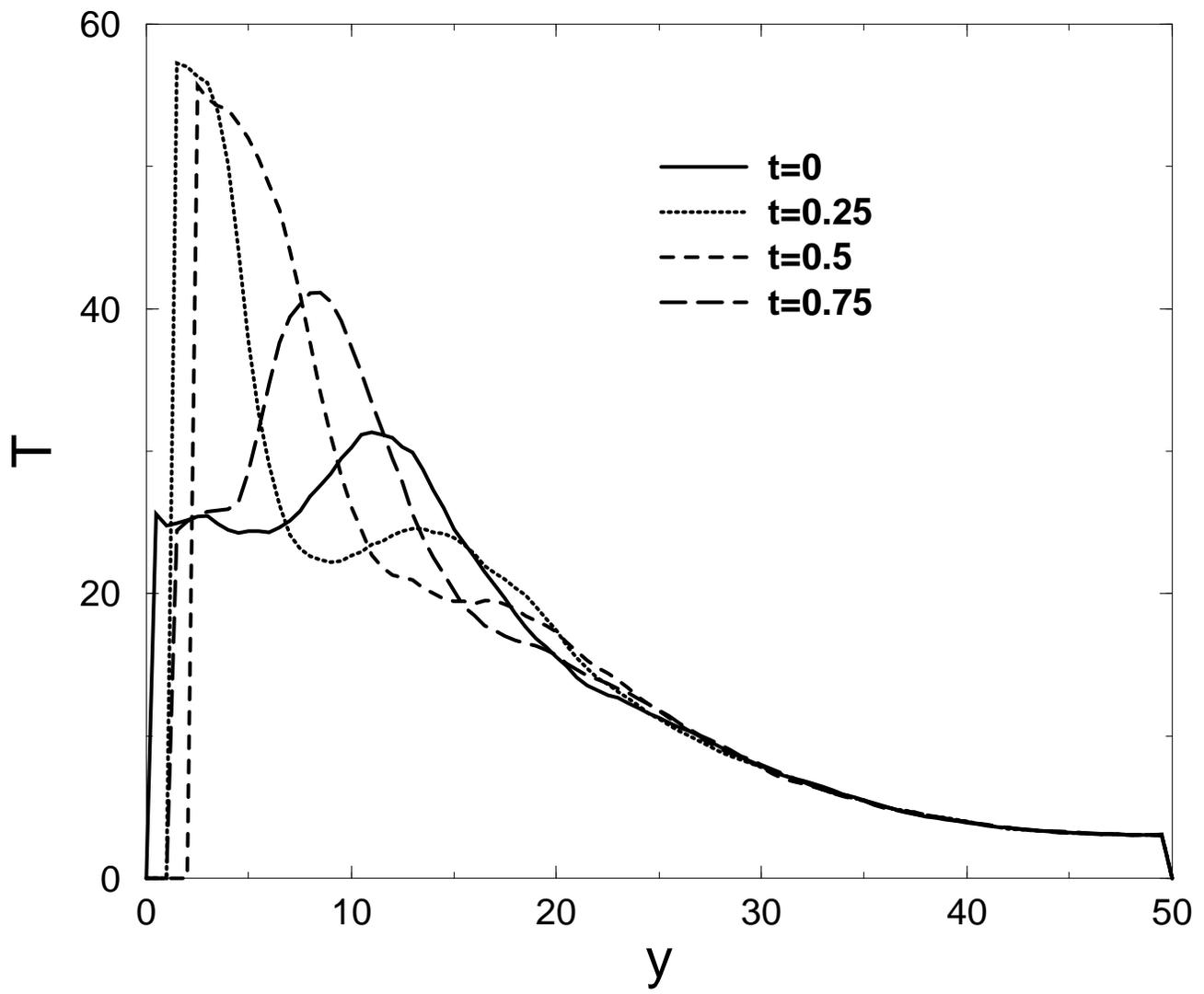
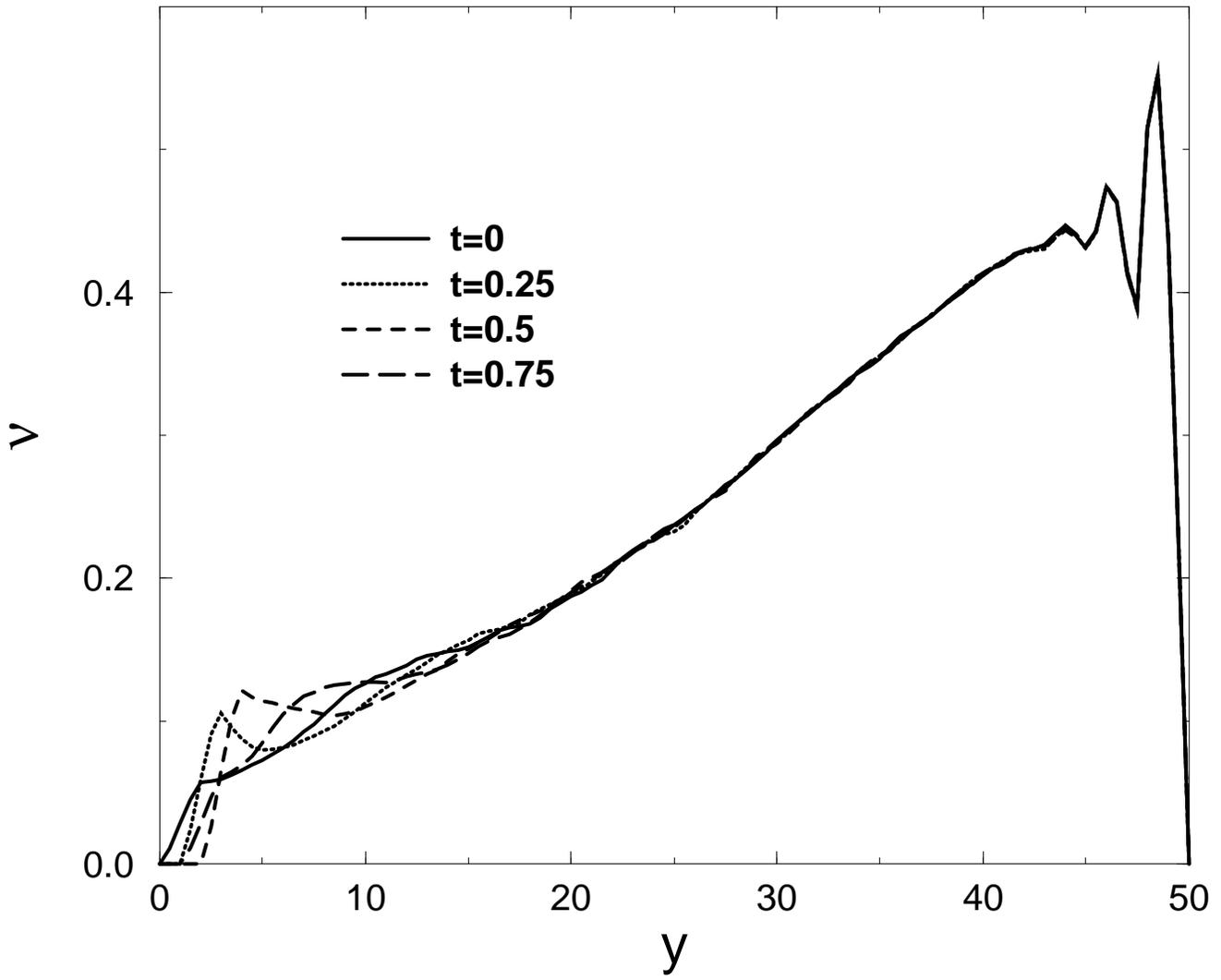
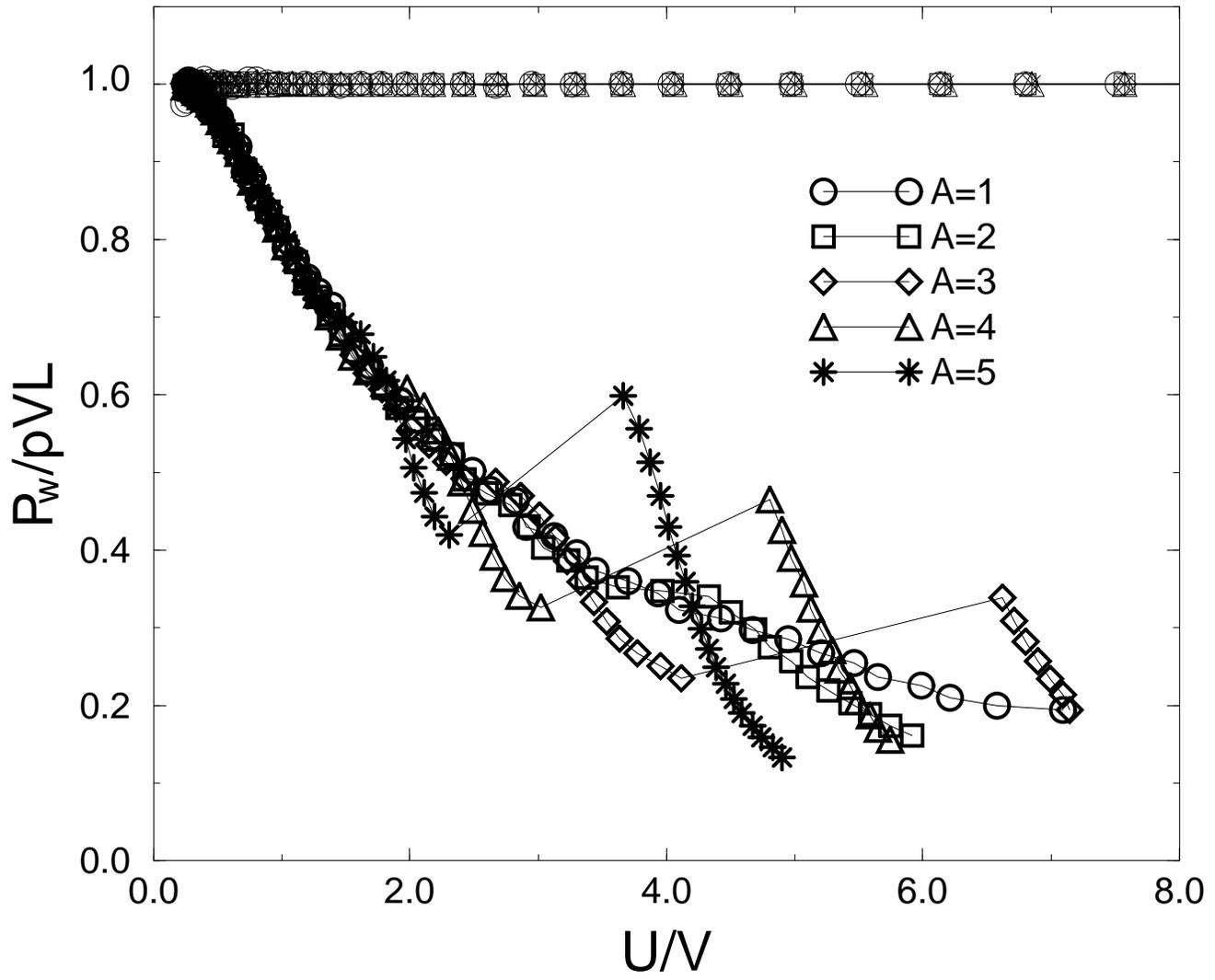
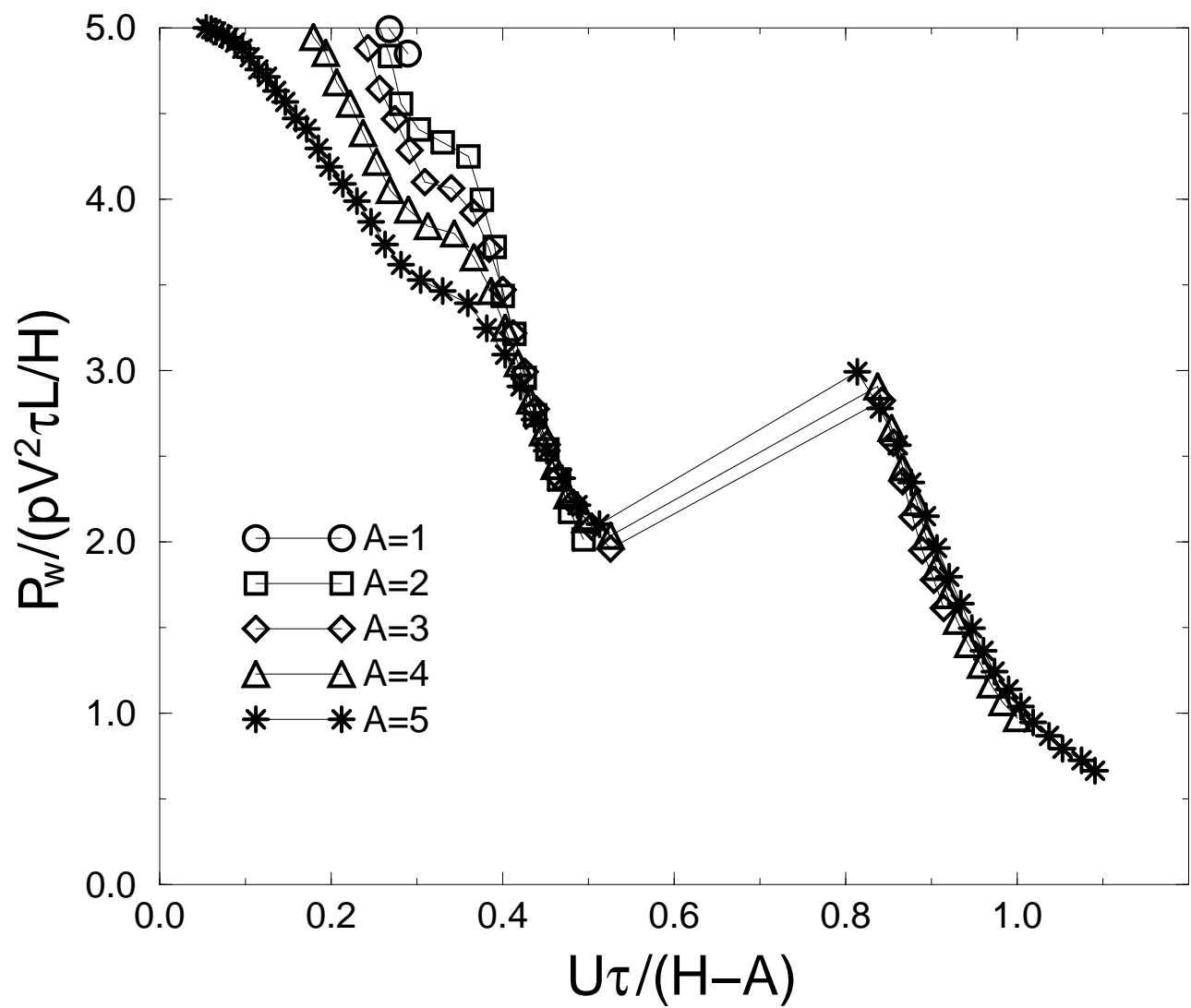
\end{document}